\documentclass[conference]{IEEEtran}
\IEEEoverridecommandlockouts
\usepackage{amsmath}
\usepackage{amsfonts}
\usepackage{amstext}
\usepackage{amssymb}
\usepackage{mathtools}

\usepackage[hyphens]{url}
\usepackage[hidelinks]{hyperref}
\hypersetup{breaklinks=true}
\usepackage{cite}

\usepackage{enumerate}
\usepackage{caption}
\usepackage{subcaption}

\usepackage{algorithm}
\usepackage{algpseudocode}
\usepackage[english]{babel}

\usepackage{epsfig}
\usepackage{xcolor}
\usepackage{color}
\usepackage{graphicx}
\usepackage{lipsum}
\usepackage{pgfplots}
\pgfplotsset{compat=newest}
\usepackage{tikz}
\usetikzlibrary{plotmarks}
\usetikzlibrary{arrows.meta}
\usetikzlibrary{shapes,arrows}
\usetikzlibrary{intersections}
\usepgfplotslibrary{patchplots}

\newtheorem{theorem}{Theorem}
\newtheorem{lemma}{Lemma}
\newtheorem{assumption}{Assumption}

\newtheorem{definition}{Definition}

\newtheorem{remark}{Remark}

\def\ba{\mathbf{a}}

\newcommand{\BS}{\mathrm{BS}}
\newcommand{\UE}{\mathrm{UE}}

\def\ps{\mathrm{ps}}
\def\PS{\mathrm{PS}}
\def\bl{\mathcal{B}}

\DeclareMathOperator*{\minimize}{minimize\;}

\addto\captionsenglish{\renewcommand{\figurename}{Fig.}}
\usepackage[top=0.75in, bottom=1in, left=0.625in, right=0.625in]{geometry}
\def\BibTeX{{\rm B\kern-.05em{\sc i\kern-.025em b}\kern-.08em
    T\kern-.1667em\lower.7ex\hbox{E}\kern-.125emX}}
\begin{document}
\title{Beam Alignment Using Trajectory Information in Mobile Millimeter-wave Networks}
\author{\IEEEauthorblockN{Sara Khosravi\IEEEauthorrefmark{1},
Hossein S. Ghadikolaei\IEEEauthorrefmark{3}, Jens Zander\IEEEauthorrefmark{1}, and Marina  Petrova \IEEEauthorrefmark{1}\IEEEauthorrefmark{2}}\\
\IEEEauthorblockA{ \IEEEauthorrefmark{1}School of EECS, KTH Royal Institute of the Technology, Stockholm, Sweden,\\   \IEEEauthorrefmark{2} Mobile Communications and Computing, RWTH Aachen University, Germany, \IEEEauthorrefmark{3} Ericsson Research, Sweden\\
Email: \{sarakhos, jenz, petrovam\} @kth.se, hossein.shokri.ghadikolaei@ericsson.com
}}
\maketitle

\begin{abstract}
Millimeter-wave and terahertz systems rely on beamforming/combining codebooks to determine the best beam directions during the initial access and data transmission. 
Existing approaches suffer from large codebook sizes and high beam searching overhead in the presence of mobile devices. 
To address this issue, we utilize the similarity of the channel in adjacent locations to divide the user trajectory into a set of separate regions and maintain a set of candidate beams for each region in a database. Due to the tradeoff between the number of regions and the signalling overhead, i.e., the greater number of regions results in a higher signal-to-noise ratio (SNR) but also a larger signalling overhead for the database, we
propose an optimization framework to find the minimum number of regions based on the trajectory of a mobile device. 
Using a ray tracing tool, we demonstrate that the proposed method provides high SNR while being more robust to the location information accuracy in comparison to the lookup table baseline and fixed size region baseline.
\end{abstract}


\section{Introduction}
Millimetre-wave (mmWave) is one of the important technologies of the fifth-generation (5G) cellular networks, offering a high rate due to the large availability of bandwidth and underutilized spectrum \cite{rappaport2013millimeter}. The current release of 5G uses mmWave bands between $24.25$ GHz and $52.6$ GHz, while the future releases are expected to include the Terahertz band too \cite{5GNR}.
However, higher frequency bands impose harsher propagation conditions and use large arrays of very small antennas \cite{Andrewprobingbeam}. Hence, mmWave networks require to use of large antenna arrays and employ directional communication to maintain viable received signal power. As the directional links are highly sensitive to blockage, finding and maintaining near-optimal directions, i.e. beam alignment, including for non-line-of-sight (NLoS) paths, is necessary. MmWave devices typically use codebooks of indexed analog beams to allow the user equipment (UE) to identify and feedback on good beams to the base station (BS). These codebooks will contain much more numerous and much narrower beams as higher frequencies are adopted, making the latency and beam searching overhead of the existing searching methods prohibitive. Therefore, beam alignment will become an increasingly important bottleneck.

In the current release of 5G, beam alignment is based on brute-force beam sweeping over the beam codebooks, measurements and reporting \cite{li2020beam,9502647}.
In the exhaustive beam sweeping, the BS and UE need to sweep almost all the combinations of beam pairs, which leads to a significant beam searching overhead.

The beam alignment methods that utilize side information such as the locations of the transmitter and
receiver have been explored to accelerate beam alignment.
In \cite{choi2016millimeter,va2017inverse}, the BS divides the serving area to equal and uniform location bins and stores the beam searching results in each bin in a lookup table. Given the location of the UE, the BS selects the beams from the lookup table. Although their solutions can reduce the searching overhead, lookup table-based methods have several limitations. First, the size of the lookup table increases linearly with the number of location bins which leads to the high signalling overhead for the lookup table. Second, uniform and equal bins may result in a high sensitivity to location information inaccuracy due to the GPS's limited accuracy. Furthermore, the optimal directions as the output of beam alignment are not merely a  function of the location but also depend on the environment geometry, blockage, etc.

Authors in \cite{sur2016beamspy,beamforecast} used the sparsity and the similarity of mmWave channels in adjacent locations and proposed their methods for the stationary \cite{sur2016beamspy} or indoor scenarios \cite{beamforecast}. 
Some approaches such as \cite{alkhateeb,heng2021machine} apply machine-learning tools to predict the optimal beams. However, they need an exhaustive or hierarchical search over all beam pairs between the BS and the UE for every location during the training phase, which may increase the complexity of the training phase.

In this work, we utilize the channel similarity of mmWave in adjacent locations and propose a beam alignment method based on available trajectory information. Our method has two phases: the training phase (offline measurement) and the run-time phase.  Some key features of the proposed method are summarized as follows.

 \begin{itemize}
 \item \textit{Designed probing beams:} The output of the training phase is the probing beam directions based on the trajectory information and environment geometry.  Our proposed method is based on dividing the UE trajectory into non-overlapping (and probably non-uniform) regions. One reference point is assigned to each region and the candidate beams will be measured in the reference points. The probing beams contain the candidate beam directions for each region of a specific trajectory. 
 \item \textit{Small number of measurements during the training phase:} Due to the tradeoff between the number of regions and signalling overhead, we propose an optimization framework to find the minimum number of regions. The proposed method only needs to run brute-force beam searching in the reference location in each region during the training phase. Unlike the existing works based on a lookup table, our proposed solution provides the minimum number of regions. 
 \item \textit{High SNR and less sensitive to the location accuracy:} During the run-time, our method can provide high SNR. Another interpretation of the regions is the maximum tolerance of noisy location input in the reference points. Hence, our method by defining the optimal size of regions provides higher robustness to the imperfect location information in comparison to the existing lookup table-based methods with predefined uniform small regions. 
      \item \textit{Empirical Evaluation:}  We apply a ray tracing tool with real building map data as an input. Simulation results demonstrate the effectiveness of our proposed method in the SNR along the trajectory and lower sensitivity to the location information accuracy.  
 \end{itemize}

The rest of the paper is organized as follows. We introduce the system model in Section \ref{system model}. In Section \ref{method}, we propose our method and in Section \ref{Simulation}, we present the numerical results. We conclude our work in Section \ref{conclusions}.

\emph{Notation}:
Sets, vectors, matrices, random variables and their realizations are denoted by calligraphic, bold small, bold capital, capital and lower case letters, respectively.
The set $\{1,\ldots, n\}$, for some integer $n$, is denoted by $[n]$.

\section{System And Channel Models}\label{system model}
We consider a downlink mmWave network where each BS has a uniform planar array (UPA) of $N_{\BS}$ antennas and each UE has a UPA of $N_{\UE}$ antennas. We assume the UE is moving along a trajectory. We consider a quantized trajectory with length $M$ and reference location indices $x\in[M]$. The trajectory length can be defined based on the coverage area of the BS. Analog beamforming with a single RF chain on both sides is assumed. Note that the beam alignment method in this paper can also be applied in hybrid architectures.

With constant block fading in the channel response during a coherence interval (CI), the channel matrix $\mathbf{H}\in\mathbb{C}^{N_{\text{UE}}\times N_{\text{BS}}}$ between the BS and UE in location $x\in [M]$ is \cite{akdeniz2014millimeter}:
\begin{equation*}
 \mathbf{H}_{x}=\sqrt{\frac{N_{\text{BS}}N_{\text{UE}}}{L}}\sum_{\ell=1}^{L} h_{x,\ell}\mathbf{a}(\phi^{\text{UE}}_{x,\ell},\theta^{\text{UE}}_{x,\ell}) \mathbf{a}^{H}(\phi^{\text{BS}}_{x,\ell},\theta^{\text{BS}}_{x,\ell}),
\end{equation*}
where $L$ is the number of available paths. 
Each path $\ell$ has horizontal and vertical angles of arrival (AoAs), $\phi^{\text{UE}}_{x,\ell}$, $\theta^{\text{UE}}_{x,\ell}$, and horizontal and vertical angles of departure (AoDs), \mbox{$\phi^{\text{BS}}_{x,\ell}$, $\theta^{\text{BS}}_{x,\ell}$,} respectively.
$h_{x,\ell} \sim \mathcal{N}_\mathbb{c} (0, \beta_{x,\ell})$ is the small scale fading, where $\beta_{x,\ell}$ is the channel gain. The half-wavelength array steering vector in the yz-plane can be written as
\begin{equation} \label{UPA}
\begin{split}
\mathbf{a}&(\phi^{S}_{x,\ell},\theta^{S}_{x,\ell})= \\
 &\frac{1}{\sqrt{N_{S}}} [1, ..., e^{j\pi [n_x \sin(\theta^S_{x,\ell})\sin(\phi^S_{x,\ell})+n_y \cos(\phi^S_{x,\ell})]},...]^T
\end{split}
\end{equation}
where $1 \leq n_x \leq N_x-1$, $1 \leq n_y \leq N_y-1$ and $N_x$ and $N_y$ are the number of columns and rows of the UPA and $N_{S}=N_x\times N_y$, $S\in \{\BS,\UE\}$.

The signal-to-noise ratio (SNR) in location $x$ is defined as ${p \lvert\mathbf{w}^{H}\mathbf{H}_x \mathbf{f}\rvert^2}/{\sigma^{2}}$,
where $p$ and $\sigma^2$ are the transmit and noise power, respectively. 
To maximize SNR, we can design beamforming vector ($\mathbf{f}\in\mathbb{C}^{N_{\text{BS}}}$) and combining vector ($\mathbf{w}\in\mathbb{C}^{N_{\text{UE}}}$) from
${\text{maximize}}_{\mathbf{f}\in \mathcal{F}, \mathbf{w}\in \mathcal{W}}
~ 	 |\mathbf{w}^H\mathbf{H}\mathbf{f}|^2 
$, 
where $\mathcal{F}$ and $\mathcal{W}$ are the transmit codebook and the receive codebooks, respectively. We define $\mathbf{f}=\mathbf{a}(\phi, \theta)$, where $(\phi, \theta)$ is the steering angle and $\mathbf{a}(.)$ is as \eqref{UPA}. The same definition is applied for $\mathbf{w}$ as well.

\begin{definition}[Path skeleton] \label{def:ps}
    The \emph{path skeleton} (PS) between the UE in location index $x$ and the BS is defined as
    \begin{equation*}
        {\PS(x) 
        := \left( \phi^{\BS}_{x,\ell}, \theta^{\BS}_{x,\ell}, 
       \phi^{\UE}_{x,\ell},\theta^{\UE}_{x,\ell}
        \right)_{\ell=1}^L}
    \end{equation*}
where $\phi^{\text{BS}}_{x,\ell}$,$ \theta^{\BS}_{x,\ell}$ and $\phi^{\text{UE}}_{x,\ell}$,$ \theta^{\UE}_{x,\ell}$ are horizontal and vertical AoD and AoA of the $\ell$-th path between the BS and a UE in location index $x$, respectively, and $L$ is the number available strong paths. Note that $\PS(x)$ is a set of steering directions of candidate beams in the location index $x$.
\end{definition}
The path skeleton can be found based on the beam-sweeping method similar to 5G. Note that due to the sparsity of the mmWave channel, the number of NLoS paths is quite small, e.g. 1 to 2 \cite{rangan2014millimeter}. Therefore, we consider the maximum size of the path skeleton equal to 3 \mbox{($L=3$)}. 

\section{Proposed Method} \label{method}
In this section, we present our method to design a database of candidate beams during the offline phase. First, we start with the problem definition. Then we present the solution.

\subsection{Problem Formulation} 
\label{Problemformulation}
\begin{figure}[!t]
    \centering
    \includegraphics[scale=0.42]{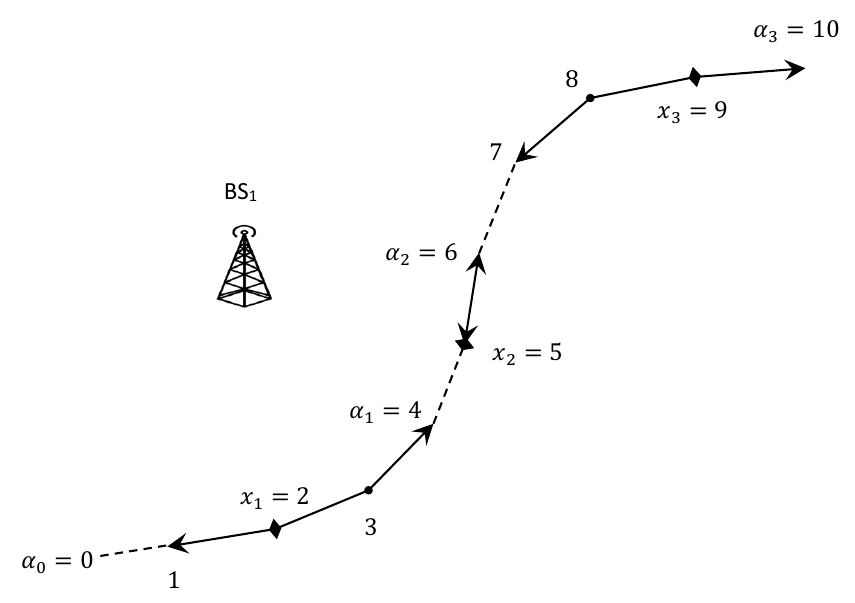}
    \caption{An example of a BS and a UE trajectory where $x\in [10]$ denotes the location index.
    The regions are $\mathcal{R}_1=\{1,2,3,4\}$, $\mathcal{R}_2=\{5,6\}$, and $\mathcal{R}_3=\{7,8,9,10\}$, with reference locations $x_1=2$, $x_2=5$, and $x_3=9$, respectively.}
    \label{fig:example}
\end{figure}

In this paper, we seek a means to decrease the beam search space and choose $\mathbf{f}$ and $\mathbf{w}$ such that the frequency of running coarse beam searching decreases while keeping UE's quality of service in terms of SNR  along the UEs trajectory. Note that our proposed method is an ad-hoc method that does not necessarily yield the optimal SNR. However, as we will show in the numerical results, it provides high SNR with less beam searching overhead along the trajectory. 
We use the correlation of path skeletons in adjacent locations and divide the trajectory into $K$ separate regions. The region $\mathcal{R}_k$ is defined as
    \begin{equation*}
        \mathcal{R}_k := \{\alpha_{k-1}+1,\ldots, \alpha_k\}   \quad k\in[K],
    \end{equation*}
where $\alpha_k\in[M]$ denote the location index of the end of the region $k$ and $0=\alpha_0\leq\alpha_1\leq\cdots\leq\alpha_K=M$.
 For region $\mathcal{R}_k$, the path skeleton of one and only one reference point $x_k\in \mathcal{R}_k\cup\{\alpha_{k-1}\}$ has been measured, and all the locations in the region use the strongest path (main beam) of the path skeleton of $x_k$ as the transmission beam. Hence, the candidate beams with steering directions $\text{PS}(x_k)$ for each region $\mathcal{R}_k$ are stored in the database. 
You can find an illustration in \mbox{Fig. \ref{fig:example}}.
There is a tradeoff between the number of regions and the signalling overhead, i.e, the number of running brute-force beam sweeping along the trajectory. The higher number of regions leads to higher SNR but also a large signalling overhead and updating of the database.
Hence, we define an optimization problem to determine the regions and reference points. To this end, first, the reference points are determined. Then, the regions are defined based on the reference points and their measured path skeleton as
\begin{subequations}
\begin{IEEEeqnarray}{rCl}
    \minimize_{\substack{
    \alpha_1,\ldots, \alpha_{K-1},\\
    x_1,\ldots,x_K}} &&~~ K \label{eqn:NonUGP} \\
    \mathrm{s.t.} && ~~
    \Pr\left\{d(x, x_k) \leq \gamma \mid \left(\ps(x_k)\right)_{k=1}^K\right\} \leq \epsilon, \nonumber\\
    &&~~\quad\forall x\in\mathcal{R}_k, 
    \quad\forall k\in[K],
    \quad\forall \left(\ps(x_k)\right)_{k=1}^K 
    \IEEEeqnarraynumspace\label{eqn:NonUGP dist} \\
    &&~~ \alpha_1\leq\cdots\leq\alpha_{K-1} \\
    &&~~ \alpha_k \in [M],
    \qquad \forall k\in[K-1] \\
    && ~~x_k \in \mathcal{R}_k \cup \{\alpha_{k-1}\},
    \qquad \forall k\in[K], \label{eqn:xkinRk}
\end{IEEEeqnarray}
\end{subequations}
where $d(x,y)$, for some $x,y\in[M]$, is defined as \eqref{d} on the top of next page.
\begin{figure*}[!t]
\begin{IEEEeqnarray}{rCl} \label{d}
 d(x,y):= 
    \sum_{\ell=1}^L \left\lvert\ba^H (\phi^{\text{UE}}_{y,\ell},\theta^{\text{UE}}_{y,\ell})\ba(\phi^{\text{UE}}_{x,\ell},\theta^{\text{UE}}_{x,\ell})\right\rvert .\left\lvert \ba^H(\phi^{\text{BS}}_{y,\ell},\theta^{\text{BS}}_{y,\ell}) \ba(\phi^{\text{BS}}_{x,\ell},\theta^{\text{BS}}_{x,\ell})\right\rvert, \IEEEeqnarraynumspace
	\end{IEEEeqnarray}
\hrulefill
\vspace*{4pt}
\end{figure*}
$\mathcal{R}_k = \{\alpha_{k-1}+1,\ldots, \alpha_k\}$ with $\alpha_0=0, \alpha_K=M$;
In \eqref{eqn:NonUGP dist}, we use pre-determined parameters $\gamma$ and $\epsilon$ to measure if the path skeleton at reference location $x_k$ is valid for any location $x\in\mathcal{R}_k$ for any possible set of path skeleton of the reference points.
The higher value of $d(\cdot,\cdot)$ means that the selected angles are close to the path skeleton in location index $x$.
Note that constraint \eqref{eqn:xkinRk} is equivalent to 
\begin{equation} \label{eqn:xk=a-1ORak}
    x_k \in \{\alpha_{k-1}, \alpha_k\},
    \qquad \forall k\in[K]
\end{equation}
because if $\alpha_{k-1} < x_k < \alpha_{k}$, we can divide $\mathcal{R}_k$ to $\mathcal{R}'_k$ from $\alpha_{k-1}$ into $x_k$ and $\mathcal{R}''_k$ from $x_k+1$ to $\alpha_k$.
In this case, each new region satisfies \eqref{eqn:xk=a-1ORak} without changing the other part of the problem.

Each reference point is determined based on the path skeleton of the previous reference points.
Note that the reference points are not necessarily chosen in ascending order.
Hence, we define $\tilde{x}_1,\ldots,\tilde{x}_K$ as a permutation of $x_1,\ldots,x_K$ such that $\tilde{x}_i$ is chosen as a reference point before $\tilde{x}_j$ for any $i<j$.
Precisely, for choosing the $k$-th reference point $\tilde{x}_k$, the set of previous reference points $\mathcal{X}_k:=\{\tilde{x}_1,\ldots,\tilde{x}_{k-1}\}\subseteq[M]$ and their path skeletons $\{\PS(z) \colon z\in\mathcal{X}_k\}$ are known:
\begin{equation} \label{eqn:refPointSelection}
    \begin{cases}
        \left(\tilde{x}_q, \PS(\tilde{x}_q)\right)_{q=1}^{k-1} 
        \mapsto \tilde{x}_k\in[M], \\
        \mathcal{X}_{k+1} = \{\tilde{x}_k\} \cup \mathcal{X}_k.
    \end{cases}
\end{equation}

In the end, the regions $\mathcal{R}_k$ are determined after evaluating all the reference points:
\begin{equation} \label{eqn:regions}
    \left(x_k\right)_{k=1}^K 
    \mapsto
    \left(\alpha_k\right)_{k=1}^{K-1}.
\end{equation}
Therefore, the optimal regions $\mathcal{R}^*_k$ and the corresponding reference points $x^*_k$, for $k\in[K]$, are determined.

\subsection{Proposed Solution} \label{Proposed Method}
Since solving \eqref{eqn:NonUGP} is difficult in practice, we assume a Markov property to make it easier.

\begin{assumption} \label{ass:Markov}
    For any $x \leq x' \leq x''\in[M]$, we have the Markov chain
    \begin{equation*}
        \PS(x)\to\PS(x')\to\PS(x'').
    \end{equation*}
\end{assumption}

We define blocks and their state in the following definition.
\begin{definition}[Blocks and states] \label{def:block}
    The block $\mathcal{B}(x_l,x_h)$, for $0 \leq x_l \leq x_h \leq M$, is a number of adjacent location indices $\{x_l+1,\ldots,x_h\}$.
    According to the measurement of the path skeleton at the start and the end of the block, we define three different blocks:
    \begin{itemize}
        \item \emph{Type 1}: A block whose path skeleton at the beginning and the end of the block is measured.
        
        \item \emph{Type 2}: A block whose path skeleton at the beginning of the block is measured.
        
        \item \emph{Type 3}: A block whose path skeleton at the end of the block is measured.
    \end{itemize}
    The state of the block $\mathcal{B}(x_l,x_h)$ is defined as
    \begin{itemize}
        \item $S := (x_l, x_h, \PS(x_l), \PS(x_h))$ for Type 1 blocks,
        \item $S := (x_l, x_h, \PS(x_l))$ for Type 2 blocks,
        \item $S := (x_l, x_h, \PS(x_h))$ for Type 3 blocks,
    \end{itemize}
\end{definition}
Note that, the whole trajectory is also a block with $x_l=0$ and $x_h=M$.
We generalize \eqref{eqn:NonUGP} and define the value of a block as the minimum number of reference points in the block.
\begin{definition}[Value of the block] \label{def:v(s)}
    For any block $\bl(x_l,x_h)$ with a given state $s$, we define the value of the block, as
\begin{subequations}
\begin{IEEEeqnarray}{rCl}
        \minimize_{\substack{
        \alpha_1,\ldots, \alpha_{K-1},\\
        x_1,\ldots,x_K}} && ~~K \label{eqn:value}\\
        \mathrm{s.t.} && ~~
      \Pr\left \{d(x, x_k) \leq \gamma \mid (\ps(x_k))_{k=1}^K\right\} \leq \epsilon, \nonumber\\
        &&\qquad\forall x\in\mathcal{R}_k, 
        \quad\forall k\in[K],
        \quad\forall (\ps(x_k))_{k=1}^K \IEEEeqnarraynumspace
        \label{eqn:d>gamma Type1}\\
        &&~~ \alpha_1\leq\cdots\leq\alpha_{K-1} \\
        && ~~\alpha_k \in \bl(x_l,x_h),
        \qquad \forall k\in[K-1] \\
        && ~~x_k \in \{\alpha_{k-1}, \alpha_k\},
        \qquad \forall k\in[K], \label{eqn:xk Type}
    \end{IEEEeqnarray}
\end{subequations}    
    where $d(\cdot,\cdot)$ is defined in \eqref{d};
    and $\mathcal{R}_k = \{\alpha_{k-1}+1,\ldots, \alpha_k\}$ with $\alpha_0=x_l, \alpha_K=x_h$.
    The reference points $x_k$ and regions $\mathcal{R}_k$ are selected based on \eqref{eqn:refPointSelection} and \eqref{eqn:regions}, respectively.
    
    We denote the value of the block in \eqref{eqn:value} by $v(s)$, $v'(s)$, or $v''(s)$, for Type 1, Type 2, or Type 3 blocks, respectively.
\end{definition}

The next lemma gives a recursive solution for the value of Type 1 blocks based on the value of smaller Type 1 blocks.
\begin{lemma} \label{lmm:OptVal1}
    Having Assumption \ref{ass:Markov}, for a Type 1 block $\mathcal{B}=\bl(x_l,x_h)$ given the state $s$, we have that
    \begin{equation} \label{eqn:v=0}
        v(s) = 0
    \end{equation}
    if for some $\alpha\in\mathcal{B}$ we have
    \begin{equation} \label{eqn:alphaType1}
        \begin{cases}
            \Pr\left\{d(x_l,x) \leq \gamma \mid s\right\} \leq \epsilon, 
            \quad\forall x\in\{x_l+1,\ldots,\alpha\}, \\
            \Pr\left\{d(x,x_h) \leq \gamma \mid s\right\} \leq \epsilon,
            \quad\forall x\in\{\alpha+1,\ldots, x_h\}.
        \end{cases}        
    \end{equation}
    In this case, the regions are $\mathcal{R}_1 = \bl(x_l,\alpha)$ and $\mathcal{R}_2 = \bl(\alpha, x_h)$.
    Otherwise,
    \begin{equation} \label{eqn:v=recV}
        v(s)
        = \minimize_{x\in\mathcal{B}\setminus\{x_h\}} 1 + v(S_1) + v(S_2),
    \end{equation}
    where $S_1$ and $S_2$ are the states of the blocks $\mathcal{B}_1=\bl(x_l,x)$ and $\mathcal{B}_2=\bl(x,x_h)$, respectively.
    Further, in \eqref{eqn:v=0}, there will be no new reference point inside the block, while in \eqref{eqn:v=recV}, the minimizer will be the next reference point inside the block.
\end{lemma}

\begin{IEEEproof}
    See Appendix \ref{sec:prf:lmm:OptVal1}.
\end{IEEEproof}

The next lemma gives a recursive solution for the value of a Type 2 block, based on the values of smaller Type 1 and Type 2 blocks.
Similar equations are valid for Type 3 blocks, based on the values of smaller Type 1 and Type 3 blocks.
\begin{lemma} \label{lmm:OptValp}
    Having Assumption \ref{ass:Markov}, for a Type 2 block $\mathcal{B}=\bl(x_l,x_h)$ given the state $s$, we have that
    \begin{equation} \label{eqn:v'=0}
        v'(s) = 0
    \end{equation}
    if we have
    \begin{equation} \label{eqn:alphaType23}
        \Pr\left\{d(x_l,x) \leq \gamma \mid s\right\} \leq \epsilon, 
        \qquad\forall x\in\mathcal{B}.
    \end{equation}
    Otherwise,
    \begin{equation} \label{eqn:v'=recVV'}
        \minimize_{x\in\mathcal{B}} 1 + v(S_1) + v'(S_2),
    \end{equation}
    where $S_1$ and $S_2$ are the states of the blocks $\mathcal{B}_1=\bl(x_l,x)$ (Type 1) and $\mathcal{B}_2=\bl(x,x_h)$ (Type 2), respectively;
    and $v(s)$ is the value of a Type 1 block with state $s$ (see Lemma \ref{lmm:OptVal1}).
    Further, in \eqref{eqn:v'=0}, there will be no new reference point inside the block, while in \eqref{eqn:v'=recVV'}, the minimizer will be the next reference point inside the block. The idea of the proof is similar to the idea of Lemma \ref{lmm:OptVal1}.
\end{lemma}


Now, we state a solution for \eqref{eqn:NonUGP}.
\begin{theorem} \label{thm:OptVal}
    Having Assumption \ref{ass:Markov}, the solution of \eqref{eqn:NonUGP} is
    \begin{equation} \label{eqn:v=recV'v''}
        \minimize_{x\in[M]} 1 + v''(S_1) + v'(S_2),
    \end{equation}
    where $S_1$ and $S_2$ are the states of the blocks $\mathcal{B}_1=\bl(1,x)$ (Type 3) and $\mathcal{B}_2=\bl(x,M)$ (Type 2), respectively;
    and $v'(s)$ and $v''(s)$ are the value of a Type 2 and Type 3 blocks with state $s$, respectively (see Lemma \ref{lmm:OptValp}).
    Further, the minimizer of the optimization will be the first reference point inside the trajectory.
    Note that $S_1$ and $S_2$ are random due to the randomness of $\PS(x)$. The idea of the proof is similar to the idea of the previous lemmas.
\end{theorem}
\begin{figure}[t]
   \centering
  \includegraphics[scale=0.075]{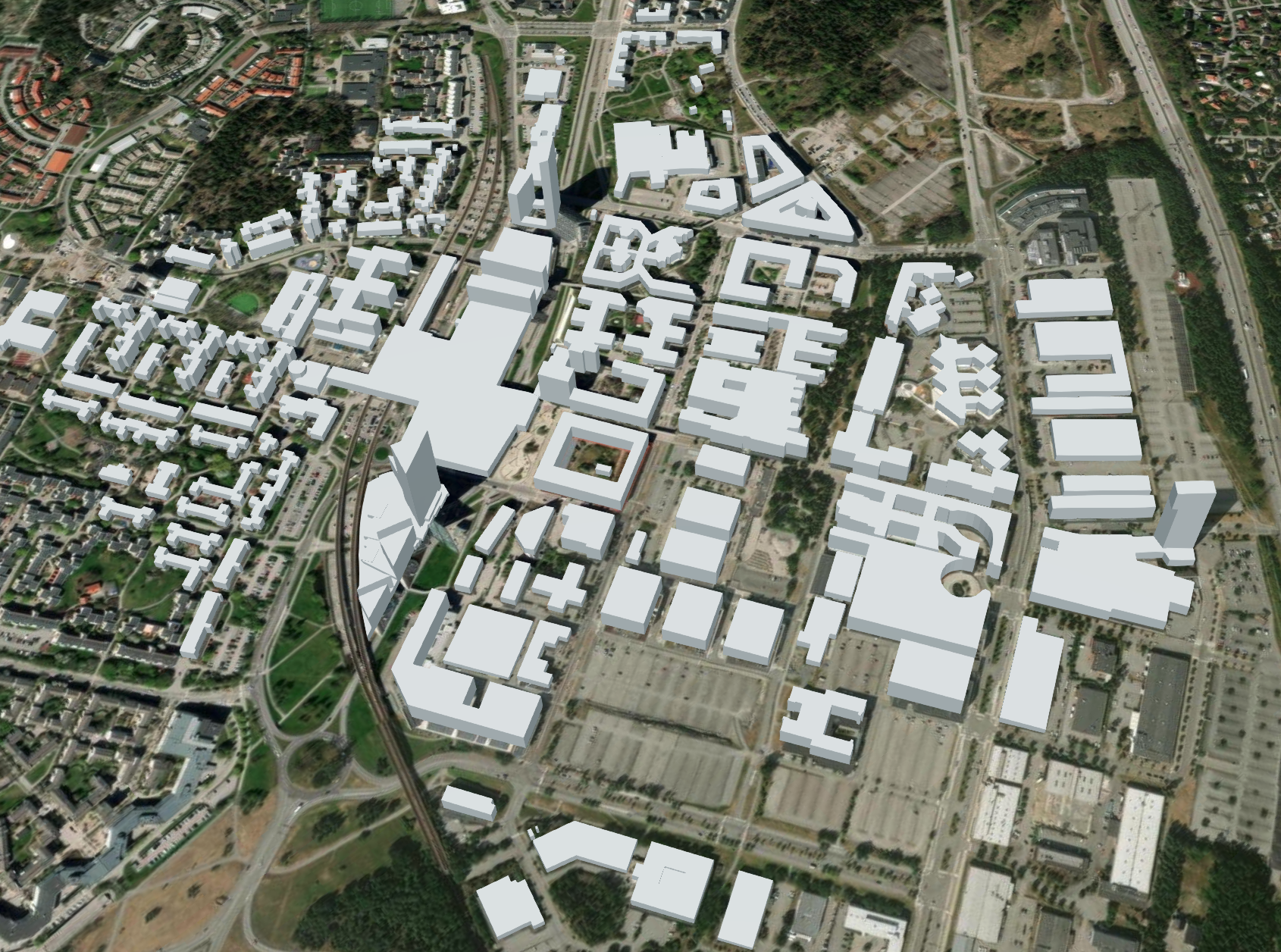}
  \caption{Simulation environment in Kista, Stockholm. Buildings are depicted in gray color. }
    \label{scenario}
\end{figure}

\begin{table}[t]
\begin{center}
\caption{Simulation parameters.}
\label{table1}
\begin{tabular}{|c|c|} 
\hline
\textbf{Parameter} & \textbf{Value} \\
\hline
\hline
$\BS$ transmit power & 10 dBm\\
 noise power ($\sigma^2$)	& -94 dBm\\
Signal bandwidth & 100 MHz \\
Carrier frequency & 28 GHz \\
BS (UE)  antennas & $8 \times 8$ ($2 \times 2$) UPA\\
$\epsilon$& 0.1\\
$\gamma$ & 0.2\\
\hline
\end{tabular}
\end{center}
\end{table}

\begin{remark}[Summary of proposed method] For a given trajectory, during the training phase, the reference locations and accordingly the regions are determined as in (14). The path skeletons are only measured in the reference locations.  The output of the training phase is a database of probing beams containing the candidate beam directions (reference path skeleton) for each region. During the run-time, the location of the UE is mapped to a specific region. The pilot signals are sent along the candidate beam directions of the corresponding region. Note that, if all the candidate beam directions of a region are blocked or weakened due to the sudden blockage, the conventional beam alignment methods can be applied. 

\end{remark}
\begin{remark} To solve \eqref{eqn:NonUGP}, we assume the joint distribution of the path skeletons in adjacent locations is available. However, in practice, we may not have
access to it. The ray tracing tool or digital twins can be applied to find the distribution statistically. Applying a machine learning tool is another approach, which we will consider in our future work.

\end{remark}

\section{Numerical Results} \label{Simulation}
\label{results}
We evaluate the performance of the proposed method in an urban environment using the ray tracing tool in the MATLAB toolbox. The output of ray tracing is the $L$ available paths between a BS and a UE in each location. 
As depicted in Fig. \ref{scenario}, we extracted the building map of Kista in Stockholm city and used it as the input for ray tracing simulation. The BSs' height is \mbox{$6$ m} and the UEs' is $1$ m.
We assumed the building material is \emph{brick} and the terrain material is \emph{concrete}. We consider the different lengths of trajectories. 
To find the solution of \eqref{eqn:v=recV'v''} for a specific trajectory, we generated $500$ trajectory samples randomly in the environment with the same length but different streets and directions.
For each trajectory, we considered one BS with a fixed position with respect to the trajectory ($10$ m above the middle of the trajectory). Hence, the location of the points of the trajectory and the BS is fixed while the environment is different. 

We consider two baselines. To have a fair comparison, we consider location-aided baselines with the training phase and run-time phase. The \textbf{baseline 1} is based on lookup table \cite{va2017inverse}. The BS divides its coverage area into a set of adjacent and equal bins with specific IDs. During the training phase (offline), the frequency of the selected beam in each bin is stored in a lookup table. During the run-time, the BS maps the UE's location to a specific bin and selects the top-ranked beam from the lookup table. 
\newline The \textbf{baseline 2} is based on fixed size regions \cite{beamforecast}. During the training phase, the regions are defined based on a fixed distance and the path skeleton of the first location of each region is chosen as the reference path skeleton. During the evaluation, the fixed size of regions is selected so that it has a similar number of regions as our proposed method. 

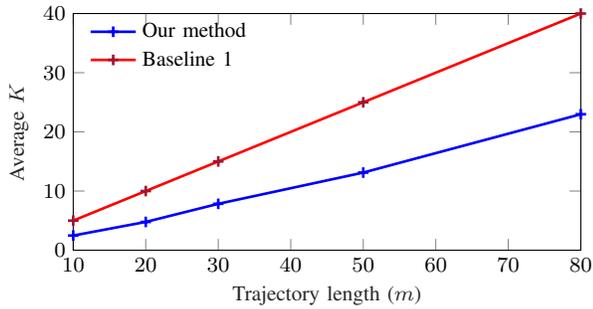
\begin{figure}[t]
	\centering
		{\footnotesize 
%
%
\definecolor{mycolor1}{rgb}{0.14902,0.14902,0.14902}%
\definecolor{mycolor2}{rgb}{0.63529,0.07843,0.18431}%
\begin{tikzpicture}

\begin{axis}[%
width=0.75\columnwidth,
height=0.35\columnwidth,
at={(0\columnwidth,0\columnwidth)},
scale only axis,
scale only axis,
xmin=10,
xmax=80,
xlabel style={font=\color{white!15!black}},
xlabel={Trajectory length ($m$)},
ymin=0,
ymax=40,
ylabel style={font=\color{white!15!black}},
ylabel={Average $K$},
axis background/.style={fill=white},
legend style={at={(0.015,0.72)}, anchor=south west, legend cell align=left, align=left, draw=white!15!black,draw=none, fill=none}
]
\addplot [color=blue, line width=1.0pt, mark size=2pt, mark=+, mark options={solid, blue}]
  table[row sep=crcr]{%
10	2.478\\
20	4.78\\
30	7.84\\
50	13.12\\
80	22.98\\
};
\addlegendentry{Our method}
\addplot [color=red, line width=1.0pt, mark size=2pt, mark=+, mark options={solid, mycolor2}]
  table[row sep=crcr]{%
10	5\\
20	10\\
30	15\\
50	25\\
80	40\\
};
\addlegendentry{Baseline 1}

\end{axis}
\end{tikzpicture}%
}
		\caption{Average number of regions for different trajectory lengths.}
		\label{average K}
\end{figure}

Fig. \ref{average K} shows the average number of $K$ over 100 trajectory samples with different lengths. Here, we consider the location bin equal to $2$ m in baseline 1. As is shown in this figure, our method has less number of regions compared to baseline 1 which leads to less number of running beam searching and signalling overhead during the training phase. Baseline 2 has the same number of regions as our method. 

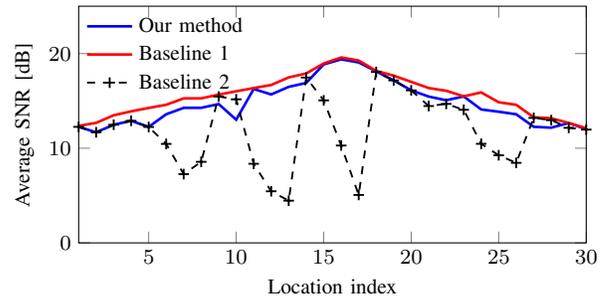
\begin{figure}[t]
	\centering
		{\footnotesize 
%
%
\begin{tikzpicture}

\begin{axis}[%
width=0.75\columnwidth,
height=0.35\columnwidth,
at={(0\columnwidth,0\columnwidth)},
scale only axis,
xmin=1,
xmax=30,
xlabel={Location index},
ymin=0,
ymax=25,
ylabel={Average SNR [dB]},
axis background/.style={fill=none},
legend style={at={(0.35,0.999)}, anchor=north east, legend cell align=left, align=left, draw=none, fill=none}
]
\addplot [color=blue, line width=1.0pt]
  table[row sep=crcr]{%
1	12.257\\
2	11.68\\
3	12.48\\
4	12.895\\
5	12.2548\\
6	13.587\\
7	14.264\\
8	14.268\\
9	14.6646\\
10	13.0141\\
11	16.257\\
12	15.68\\
13	16.48\\
14	16.895\\
15	18.8548\\
16	19.387\\
17	19.064\\
19	17.1646\\
20	16.1141\\
21	15.457\\
22	15.068\\
23	15.47\\
24	14.095\\
25	13.8548\\
26	13.587\\
27	12.264\\
28	12.168\\
29	12.6646\\
30	12.0041\\
};
\addlegendentry{Our method}

\addplot [color=red, line width=1.0pt]
  table[row sep=crcr]{%
1	12.357\\
2	12.68\\
3	13.48\\
4	13.895\\
5	14.2548\\
6	14.587\\
7	15.264\\
8	15.268\\
9	15.6646\\
10	16.0141\\
11	16.357\\
12	16.68\\
13	17.48\\
14	17.895\\
15	18.9548\\
16	19.587\\
17	19.264\\
18	18.168\\
19	17.6646\\
20	17.0141\\
21	16.357\\
22	16.068\\
23	15.48\\
24	15.895\\
25	14.8548\\
26	14.587\\
27	13.264\\
28	13.168\\
29	12.6646\\
30	12.1141\\
};
\addlegendentry{Baseline 1 }

\addplot [color=black, dashed, line width=0.7pt, mark size=2pt, mark=+, mark options={solid, black}]
  table[row sep=crcr]{%
1	12.257\\
2	11.68\\
3	12.48\\
4	12.895\\
5	12.2448\\
6	10.456\\
7	7.264\\
8	8.5645\\
9	15.4546\\
10	15.1654\\
11	8.357\\
12	5.45\\
13	4.46\\
14	17.46678\\
15	15.0548\\
16	10.287\\
17	5.064\\
18	18.05645\\
19	17.1646\\
20	16.1141\\
21	14.457\\
22	14.68\\
23	14.06\\
24	10.46678\\
25	9.2548\\
26	8.456\\
27	13.204\\
28	12.9645\\
29	12.1546\\
30	11.9654\\
};
\addlegendentry{Baseline 2}

\end{axis}
\end{tikzpicture}%
}
		\caption{Average SNR along the trajectory with perfect location of the UE.}
		\label{average SNR}
\end{figure}
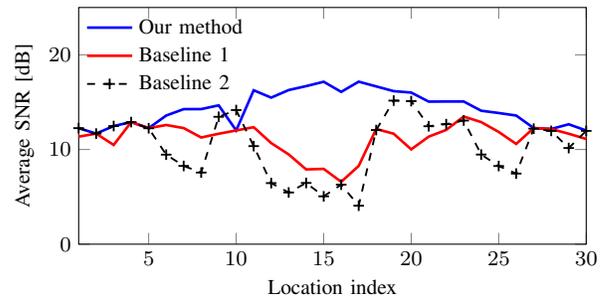
\begin{figure}[t]
	\centering
		{\footnotesize 
%
%
\begin{tikzpicture}

\begin{axis}[%
width=0.75\columnwidth,
height=0.35\columnwidth,
at={(0\columnwidth,0\columnwidth)},
scale only axis,
xmin=1,
xmax=30,
xlabel={Location index},
ymin=0,
ymax=25,
ylabel={Average SNR [dB]},
axis background/.style={fill=none},
legend style={at={(0.35,0.999)}, anchor=north east, legend cell align=left, align=left, draw=none, fill=none}
]
\addplot [color=blue, line width=1.0pt]
  table[row sep=crcr]{%
1	12.257\\
2	11.68\\
3	12.48\\
4	12.895\\
5	12.2548\\
6	13.587\\
7	14.264\\
8	14.268\\
9	14.6646\\
10	12.0141\\
11	16.257\\
12	15.48\\
13	16.28\\
14	16.695\\
15	17.1548\\
16	16.087\\
17	17.164\\
19	16.1646\\
20	16.0141\\
21	15.057\\
22	15.068\\
23	15.07\\
24	14.095\\
25	13.8548\\
26	13.587\\
27	12.264\\
28	12.168\\
29	12.6646\\
30	12.0041\\
};
\addlegendentry{Our method}

\addplot [color=red, line width=1.0pt]
  table[row sep=crcr]{%
1	11.357\\
2	11.68\\
3	10.48\\
4	12.895\\
5	12.2548\\
6	12.587\\
7	12.264\\
8	11.268\\
9	11.6646\\
10	12.0141\\
11	12.357\\
12	10.68\\
13	9.48\\
14	7.895\\
15	7.9548\\
16	6.587\\
17	8.264\\
18	12.168\\
19	11.6646\\
20	10.0141\\
21	11.357\\
22	12.068\\
23	13.48\\
24	12.895\\
25	11.8548\\
26	10.587\\
27	12.264\\
28	12.168\\
29	11.6646\\
30	11.1141\\
};
\addlegendentry{Baseline 1 }

\addplot [color=black, dashed, line width=0.7pt, mark size=2pt, mark=+, mark options={solid, black}]
  table[row sep=crcr]{%
1	12.257\\
2	11.68\\
3	12.48\\
4	12.895\\
5	12.2448\\
6	9.456\\
7	8.264\\
8	7.5645\\
9	13.4546\\
10	14.1654\\
11	10.357\\
12	6.45\\
13	5.46\\
14	6.46678\\
15	5.0548\\
16	6.287\\
17	4.064\\
18	12.05645\\
19	15.1646\\
20	15.1141\\
21	12.457\\
22	12.68\\
23	13.06\\
24	9.46678\\
25	8.2548\\
26	7.456\\
27	12.204\\
28	11.9645\\
29	10.1546\\
30	11.9654\\
};
\addlegendentry{Baseline 2}

\end{axis}
\end{tikzpicture}%
}
		\caption{Average SNR along the trajectory with noisy location of the UE.}
		\label{noisy SNR}
\end{figure}

\begin{figure}
   \centering  
  \includegraphics[scale=0.13]{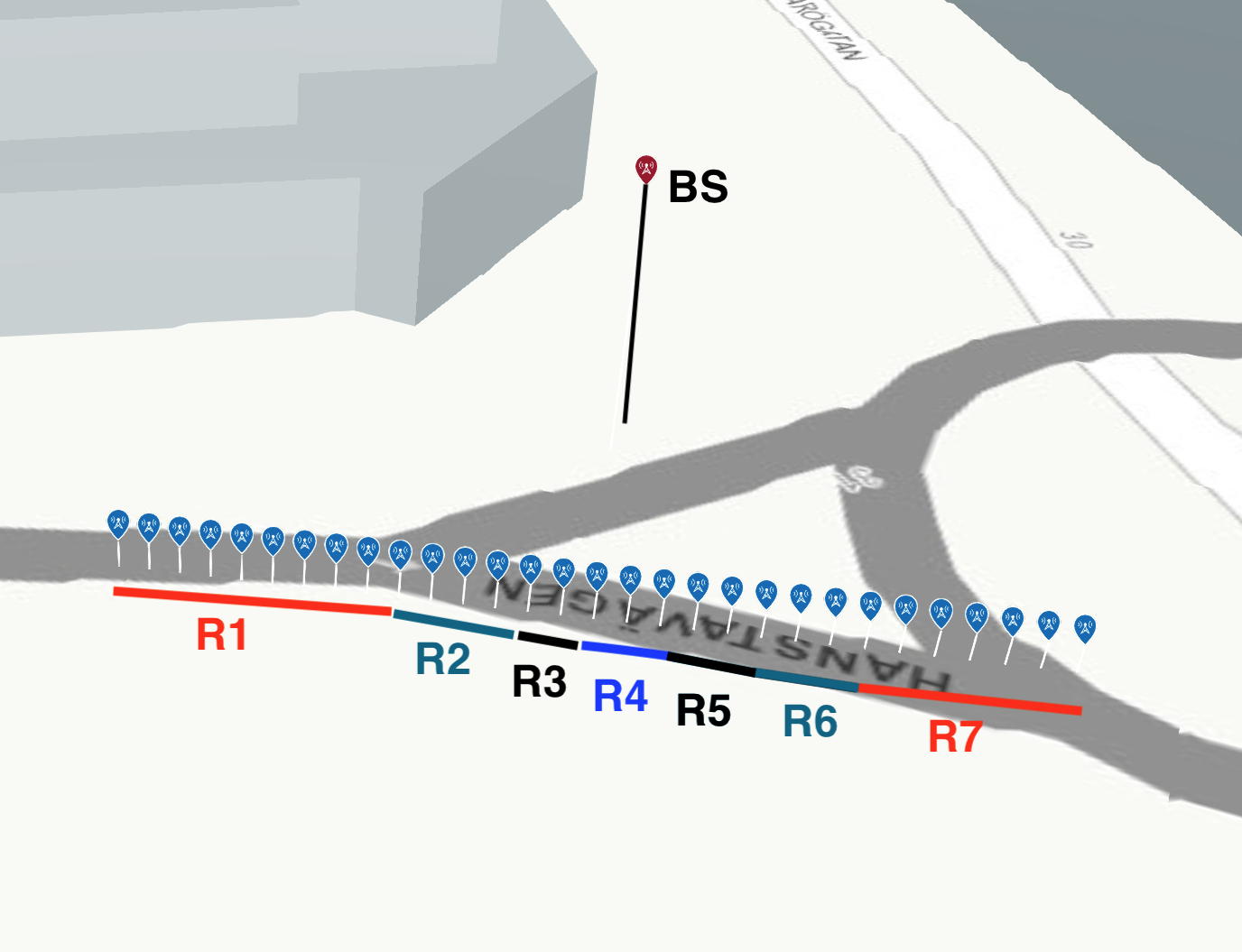}
  \caption{A trajectory sample with length 30 $m$ and $7$ regions. The blue signs show the 30 location indices every 1 $m$. }
    \label{example2}
\end{figure}

We fix the trajectory length to $30$ meters. Fig. \ref{example2}, shows the distribution of the $7$ regions. We can observe that in location indices near the BS, we have smaller regions in comparison with the beginning and the end points of the trajectory. In baseline 1, we consider the bin size equal to $2$ meter. Hence during the training phase, it needs to run the brute-force beam searching 15 times for the trajectory with a length of $30$ meters to find the optimal beam for each bin. In baseline 2, the fixed size region is about $4.5$ meter to have the same number of the regions ($K=7$) as our proposed beam alignment method. 

During the run-time phase first, we consider the perfect location information. Fig. \ref{average SNR} shows the average SNR along the trajectory. We observe that our proposed method can provide almost the same SNR value along the trajectory with about half the number of measurements during the training phase as compared to baseline 1 with higher measurements. In baseline 2 non-optimal selections of reference locations and regions cause high SNR fluctuations along the trajectory. 
\newline Next, we consider noisy location information.
Noise in GPS coordinates due to the feedback latency and user mobility causes imperfect knowledge of UE locations. Here, we model it with an addictive zero-mean Gaussian noise with a standard deviation of $2.04$ meters to the Cartesian coordination of the UE. The noise variance is selected so that the probability of noise magnitude being higher than 5 meters is less than 5$\%$( the mean GPS accuracy on smartphones is about 4.9 meters) \cite{heng2021machine}. Fig. \ref{noisy SNR} shows the SNR value in different locations of the UE along the trajectory. We can clearly observe the sensitivity of the methods to the location information accuracy is high in the middle of the trajectory (when the UE is close to the BS). Our method, however, by optimally selecting the size of the regions and reference locations is less sensitive to the imperfect location input compared with the baselines. 

\section{Conclusion} \label{conclusions}
We proposed a beam alignment method that utilizes channel similarity in adjacent locations. Our method is based on dividing the trajectory into separate regions. We propose an optimization problem to find the number of regions and proposed a solution for that. 
Simulation results verified the better performance of our method in SNR with noisy location information in comparison with the baselines. 

\Urlmuskip=0mu plus 1mu\relax
\bibliographystyle{IEEEtran}
\bibliography{IEEEabrv,ref_bib}

\begin{thebibliography}{10}
\providecommand{\url}[1]{#1}
\csname url@samestyle\endcsname
\providecommand{\newblock}{\relax}
\providecommand{\bibinfo}[2]{#2}
\providecommand{\BIBentrySTDinterwordspacing}{\spaceskip=0pt\relax}
\providecommand{\BIBentryALTinterwordstretchfactor}{4}
\providecommand{\BIBentryALTinterwordspacing}{\spaceskip=\fontdimen2\font plus
\BIBentryALTinterwordstretchfactor\fontdimen3\font minus
  \fontdimen4\font\relax}
\providecommand{\BIBforeignlanguage}[2]{{%
\expandafter\ifx\csname l@#1\endcsname\relax
\typeout{** WARNING: IEEEtran.bst: No hyphenation pattern has been}%
\typeout{** loaded for the language `#1'. Using the pattern for}%
\typeout{** the default language instead.}%
\else
\language=\csname l@#1\endcsname
\fi
#2}}
\providecommand{\BIBdecl}{\relax}
\BIBdecl

\bibitem{rappaport2013millimeter}
T.~S. Rappaport, S.~Sun, R.~Mayzus, H.~Zhao, Y.~Azar, K.~Wang, G.~N. Wong,
  J.~K. Schulz, M.~Samimi, and F.~Gutierrez~Jr, ``Millimeter wave mobile
  communications for {5G} cellular: It will work!'' \emph{IEEE Access}, vol.~1,
  no.~1, pp. 335--349, May 2013.

\bibitem{5GNR}
\emph{Study on supporting NR from 52.6 GHz to 71 GHz (Release 17)}, Document
  3GPP TR 38.808, Mar. 2021.

\bibitem{Andrewprobingbeam}
Y.~Heng, J.~Mo, and J.~G. Andrews, ``Learning probing beams for fast mmwave
  beam alignment,'' pp. 1--6, 2021.

\bibitem{li2020beam}
Y.-N.~R. Li, B.~Gao, X.~Zhang, and K.~Huang, ``Beam management in
  millimeter-wave communications for 5g and beyond,'' \emph{IEEE Access},
  vol.~8, pp. 13\,282--13\,293, 2020.

\bibitem{9502647}
Y.~Heng, J.~G. Andrews, J.~Mo, V.~Va, A.~Ali, B.~L. Ng, and J.~C. Zhang, ``Six
  key challenges for beam management in 5.5g and 6g systems,'' \emph{IEEE
  Communications Magazine}, vol.~59, no.~7, pp. 74--79, 2021.

\bibitem{choi2016millimeter}
J.~Choi, V.~Va, N.~Gonzalez-Prelcic, R.~Daniels, C.~R. Bhat, and R.~W. Heath,
  ``Millimeter-wave vehicular communication to support massive automotive
  sensing,'' \emph{IEEE Communications Magazine}, vol.~54, no.~12, pp.
  160--167, 2016.

\bibitem{va2017inverse}
V.~Va, J.~Choi, T.~Shimizu, G.~Bansal, and R.~W. Heath, ``Inverse multipath
  fingerprinting for millimeter wave v2i beam alignment,'' \emph{IEEE Trans. on
  Veh. Technol.}, vol.~67, no.~5, pp. 4042--4058, 2017.

\bibitem{sur2016beamspy}
S.~Sur, X.~Zhang, P.~Ramanathan, and R.~Chandra, ``Beamspy: Enabling robust 60
  {GHz} links under blockage.'' in \emph{USENIX NSDI}, 2016, pp. 193--206.

\bibitem{beamforecast}
A.~Zhou, X.~Zhang, and H.~Ma, ``Beam-forecast: Facilitating mobile 60 {GHz}
  networks via model-driven beam steering,'' in \emph{Pro. IEEE Conference on
  Computer Communications (INFOCOM)}, 2017, pp. 1--9.

\bibitem{alkhateeb}
Y.~Zhang, M.~Alrabeiah, and A.~Alkhateeb, ``Reinforcement learning of beam
  codebooks in millimeter wave and terahertz mimo systems,'' \emph{IEEE
  Transactions on Communications}, vol.~70, no.~2, pp. 904--919, 2022.

\bibitem{heng2021machine}
Y.~Heng and J.~G. Andrews, ``Machine learning-assisted beam alignment for
  mmwave systems,'' \emph{IEEE Transactions on Cognitive Communications and
  Networking}, vol.~7, no.~4, pp. 1142--1155, 2021.

\bibitem{akdeniz2014millimeter}
M.~R. Akdeniz, Y.~Liu, M.~K. Samimi, S.~Sun, S.~Rangan, T.~S. Rappaport, and
  E.~Erkip, ``Millimeter wave channel modeling and cellular capacity
  evaluation,'' \emph{IEEE Journal on Selected Areas in Communications},
  vol.~32, no.~6, pp. 1164--1179, Jun. 2014.

\bibitem{rangan2014millimeter}
S.~Rangan, T.~S. Rappaport, and E.~Erkip, ``Millimeter-wave cellular wireless
  networks: Potentials and challenges,'' \emph{Proceedings of the IEEE}, vol.
  102, no.~3, pp. 366--385, 2014.

\end{thebibliography}

\begin{appendices}
\section{Proof of Lemma \ref{lmm:OptVal1}} \label{sec:prf:lmm:OptVal1}
It is clear that if \eqref{eqn:alphaType1} is valid, there is no need to select any other reference point inside the block; as a result, the solution of \eqref{eqn:value} is $0$ and the regions are $\bl(x_l,\alpha)$ and $\bl(\alpha, x_h)$.

However, if \eqref{eqn:alphaType1} is not satisfied, it is required to have at least one reference point inside the block due to the constraint \eqref{eqn:xk Type}.
Let $\tilde{x}_1\in\{x_l+1,x_h-1\}$ be the next reference point with the path skeleton $\PS(x)$, which is random given $s$.
The other reference points are $\mathcal{X}'=\{\tilde{x}_2, \ldots, \tilde{x}_K\}$.
Consider two blocks $\bl_1=\bl(x_l,\tilde{x}_1)$ and $\bl_2=\bl(\tilde{x}_2,x_h)$.
Define 
\begin{IEEEeqnarray*}{l}
    \mathcal{X}_1
    = \{x_1^{(1)},\ldots,x_{K_1}^{(1)}\}
    := \mathcal{X}' \cap \bl_1, \\
    \mathcal{X}_2
    = \{x_1^{(2)},\ldots,x_{K_2}^{(2)}\}
    := \mathcal{X}' \cap \bl_2,
\end{IEEEeqnarray*}
where $K_1$ and $K_2$ are the number of reference points inside the blocks $\bl_1$ and $\bl_2$, respectively.
Therefore, 
\begin{equation} \label{eqn:K=k1+k2+1}
    K = K_1 + K_2 + 1,
\end{equation}
where $1$ is added because of $\tilde{x}_1$.
We also define the non-overlapping regions $\mathcal{R}_1^{1},\ldots,\mathcal{R}_{K_1}^{1}$ and $\mathcal{R}_1^{2},\ldots,\mathcal{R}_{K_2}^{2}$ as the optimal regions inside $\bl_1$ and $\bl_2$, respectively (Note that we can do so because of \eqref{eqn:xk Type} stating that the reference point of any region is at the beginning or the end of the region).
Therefore,
\begin{equation*}
    \left\{\mathcal{R}_k\right\}_{k=1}^K 
    = \left\{\mathcal{R}_k^{1}\right\}_{k=1}^{K_1}
    \cup \left\{\mathcal{R}_k^{2}\right\}_{k=1}^{K_2},
\end{equation*}
where $\mathcal{R}_k$ was defined in Definition \ref{def:v(s)}.

Hence, we can write the optimization problem \eqref{eqn:value} as follows
\begin{IEEEeqnarray}{rCl}
    \min_{\tilde{x}_1}
    && \:\big[1 +
    \minimize_{\substack{
    \alpha_1,\ldots, \alpha_{K_1-1},\\
    x_1^{(1)},\ldots,x^{(1)}_{K-1}}} 
    {K_1}+ \minimize_{\substack{
    \alpha_{K_1+1},\ldots, \alpha_{K_1+K_2},\\
    x_1^{(2)},\ldots,x^{(2)}_{K-1}}}{K_2}\big] \label{eqn:E[K|s]:prf}\\
    \text{s.t.:}\: && 
    \Pr\left\{d(x, x_k^{(1)}) \leq \gamma \mid \left(\ps(x_k^{(1)})\right)_{k=1}^{K_1}, 
    \ps(x_l), \ps(\tilde{x}_1)\right\} \nonumber\\
    &&\qquad\qquad\leq \epsilon,
    \qquad\forall x\in\mathcal{R}_k^{1}, 
    \forall k\in[K_1], \nonumber\\
    &&\qquad\forall \left(\ps(x_k^{(1)})\right)_{k=1}^{K_1}, 
    \ps(x_l), \ps(\tilde{x}_1)
    \label{eqn:d>gamma Type1 B1}\\
    &&\Pr\left\{d(x, x_k^{(2)}) \leq \gamma \mid \left(\ps(x_k^{(2)})\right)_{k=1}^{K_2}, 
    \ps(x_h), \ps(\tilde{x}_1)\right\} \nonumber\\
    &&\qquad\qquad\leq \epsilon, 
    \qquad\forall x\in\mathcal{R}_k^{2}, 
    \forall k\in[K_2], \nonumber\\
    &&\qquad\forall \left(\ps(x_k^{(2)})\right)_{k=1}^{K_2}, 
    \ps(x_h), \ps(\tilde{x}_1)
    \label{eqn:d>gamma Type1 B2}\\
    && \alpha_1\leq\cdots\leq\alpha_{K-1} \\
    && \alpha_k \in \bl(x_l,\tilde{x}_1),
    \qquad \forall k\in[K_1-1] \\
    && \alpha_k \in \bl(\tilde{x}_1,x_h),
    \qquad \forall k\in\{K_1+1,\ldots,K-1\} \\
    && x_k \in \{\alpha_{k-1}, \alpha_k\},
    \qquad \forall k\in[K], \label{eqn:xk Type1}
\end{IEEEeqnarray}
Thus, optimization problem \eqref{eqn:E[K|s]:prf} is decomposed to two independent optimization problems and \eqref{eqn:v=recV} follows, and the lemma is proved.
\eqref{eqn:d>gamma Type1 B1} and \eqref{eqn:d>gamma Type1 B2} follow from \eqref{eqn:d>gamma Type1} and Assumption \ref{ass:Markov} because for all $z_1\in\bl_1$ and $z_2\in\bl_2$, $\PS(z_1)$ is independent of $\PS(z_2)$ given $\PS(\tilde{x}_1)$.
\end{appendices}
\end{document}